\documentclass[aps,prb,twocolumn,floatfix,amsmath,amssymb]{revtex4}
\usepackage[final]{graphicx}
\usepackage{bm}
\usepackage{float}
\usepackage{afterpage}

\begin{document}

\title{Hard-core bosons in one-dimensional interacting topological bands}

\author{Huaiming Guo}
\affiliation{Department of Physics, Beihang University, Beijing, 100191, China}

\begin{abstract}
We study the hard-core bosons in one-dimensional (1D) interacting topological bands at different filling factors using exact diagonalization. At the filling factor $\nu=1$ and in the presence of on-site Hubbard interaction, we find no sign of the existence of the bosonic topological phase, which is in contrast to the fermionic case. Instead by studying the momentum distribution and the condensate fraction we find a superfluid (SF) to Mott-insulator transition driven by the Hubbard interaction. At the filling factor $\nu=1/3$ and in the presence of longer-ranged interactions, we identify the bosonic fractional topological phase (FTP) whose ground-states are characterized by the three-fold degeneracy and quantized total Berry phase, which is very similar to the fermionic case. Finally we discuss the reason of the different behaviors of hard-core bosons at different filling factors by mapping them to spinless fermions. Our these results can be realized in cold-atom experiments.
\end{abstract}


\maketitle
\section{Introduction}
Recently topological insulators (TIs) attract intense theoretical and experimental studies\cite{kane1,kane2}. Till now many materials are found to be TIs. The properties of non-interacting TIs have been well understood and some of their important properties are verified by experiments \cite{moore1,hasan1,xlqi1}. Meanwhile the effects of interactions in TIs begin to be explored numerically and analytically\cite{int0,int1,int2,int3,int4,int5,int6}. At the mean-field level, the interaction can be decoupled to generate spin-orbit coupling and topological Mott insulator is realized\cite{int0}. Numerical simulations using different methods obtain consistent results\cite{int1,int2,int3,int4,int5}. The interacting topological invariant is developed using Green's function and simplified formula is proposed in terms of the Green's function at zero frequency or in the presence of inversion symmetry\cite{green1,green2,green3}. The effects of interactions on the topological classification of free fermion systems are also studied\cite{class1,class2}.

By analogy with TIs, models that exhibit nearly flatband with non-trivial topology are constructed in different systems, in which fractional Chern insulators (FCIs) may be realized in the absence of external magnetic fields \cite{flat1,flat2,flat3,flat4,flat5,flat6,flat7,flat8,fqh1,fqh2,fqh3,fqh4}. The phase is characterized by a multi-fold degenerate ground-states with quantized total Chern number. By combining the two copies of FCIs formed by spin up and down electrons, fractional TIs with time-reversal symmetry can be constructed, which will be another new quantum state of matter.

In real materials, the properties are usually exhibited by electrons which are fermions, so most of the above studies are for fermions. It is also interesting to ask whether there exist similar topological phases in bosonic systems. The studies in two dimensions have appeared and the properties of hard-core bosons in the topological bands are investigated \cite{int1,fqh4}.
In this paper, based on our study on the 1D interacting fermionic model \cite{my1,my2}, we study the behavior of hard-core bosons in 1D topological bands using exact diagonalization.

We study the cases of the filling factor $\nu=1$ and $\nu=1/3$.
For the case of $\nu=1$, we consider on-site Hubbard interaction. By calculating the energies of the lowest states, the Berry phase and the fidelity metric of the ground-states, we find no sign of the existence of the bosonic topological phase. We further calculate the momentum distribution and the condensate fraction and find a SF to Mott-insulator transition driven by the Hubbard interaction. For the case of $\nu=1/3$, we consider nearest-neighboring (NN) and next-nearest-neighboring (NNN) interactions. We identify the bosonic FTP whose ground-states are characterized by the three-fold degeneracy and quantized total Berry phase. The obtained phase diagram is very similar to that of the corresponding fermionic system except the different critical values. Finally we discuss the reason of the different behaviors of hard-core bosons at different filling factors.
\section{The model}

Our starting point is the 1D interacting tight-binding model filled with hard-core bosons \cite{my1},
\begin{eqnarray}\label{eq1}
H&=&\sum_{i} (M+2B)\Psi _{i}^{\dagger }\sigma _{z}\Psi_{i}-\sum_{i,\hat{x}} B\Psi _{i}^{\dagger }\sigma _{z}\Psi_{i+\hat{x}} \\ \nonumber
&-&\sum_{i,\hat{x}} sgn(\hat{x})iA\Psi _{i}^{\dagger }\sigma _{x}\Psi_{i+\hat{x}}+U\sum_{i}
n_{i}^{c}n_{i}^{d}
\end{eqnarray}
where $\sigma_x$, $\sigma_z$ are Pauli matrices, $\Psi_{i}=(c_{i},d_{i})^{T}$ with $c_{i}$($d_{i}$) hard-core boson annihilating operator at the site ${\bf r}_i$ and $n_{i}^{c} (n_{i}^{d})$ is the number operator of the orbit $c (d)$. In the fermionic version of the non-interacting model ($U=0$), depending on the values of the parameters $A$, $B$ and $M$ the system can be a trivial insulator or a non-trivial insulator at half-filling. Though like spinless fermion the occuping number of hard-core bosons is $0$ or $1$ per orbit on each site, the hard-core bosons obey commutation relation $[c_i,c_j^{\dagger}] ([d_i,d_j^{\dagger}])=0$ at sites $i\neq j$ but anticommutation $\{c_i,c_i^{\dagger}\} (\{d_i,d_i^{\dagger}\})=1$ on sole site $i$, which makes the hard-core bosons exhibit different properties from the fermions. In the following calculations, we focus on the parameters' region where the Hamiltonian Eq.(\ref{eq1}) at $U=0$ has non-trivial fermionic topological phase and study the properties of hard-core bosons in the interacting topological bands at different fillings.

\begin{figure}[tbp]
\includegraphics[width=8cm]{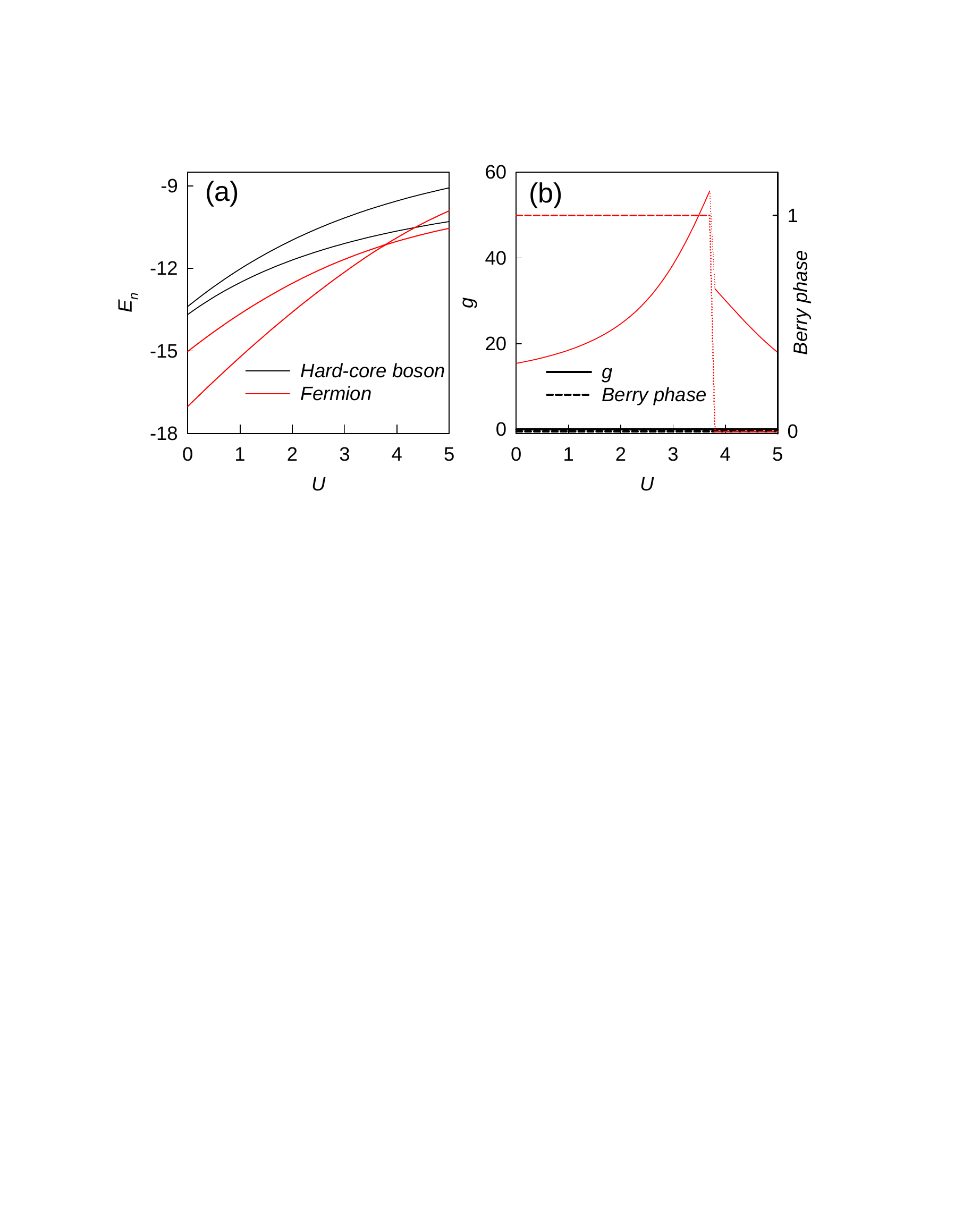}
\caption{(Color online) (a) The energies of the ground- and first-excited states vs $U$. (b) The Berry phase and the fidelity metric vs $U$. The results in bosonic (black) and fermionic (red) systems are compared. The parameters are $A=B=1$, $M=-1$ and the system size is $L=8$.}
\label{fig1}
\end{figure}
\begin{figure}[tbp]
\includegraphics[width=8cm]{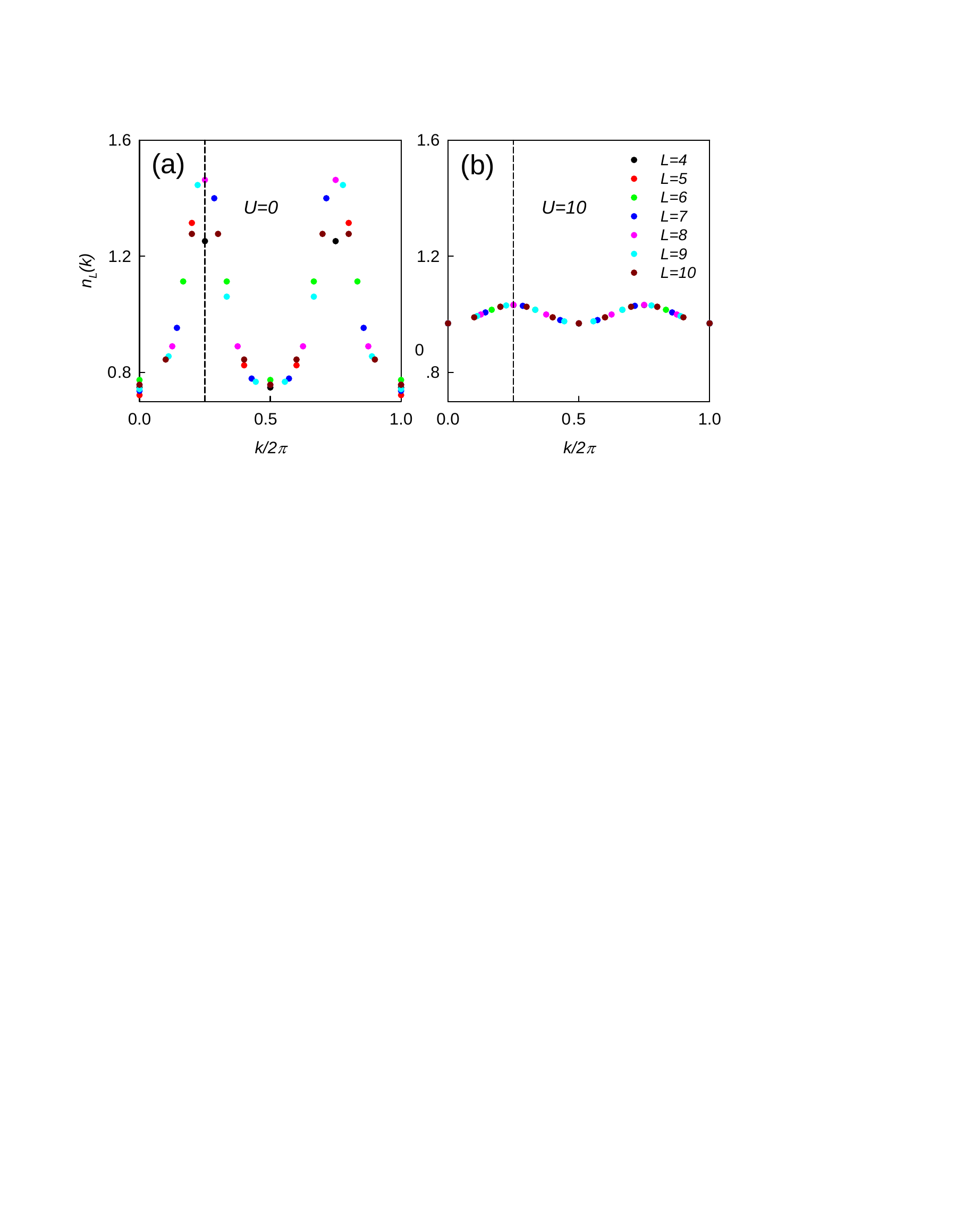}
\caption{(Color online)The momentum distributions at (a) $U=0$ and (b) $U=10$ on different sizes. The parameters are the same with those in Fig.\ref{fig1}.}
\label{fig2}
\end{figure}
%

%
\section{The filling factor $\nu=1$}
We first study the case of the filling factor $\nu=1$ (we denote the number of particles as $N_p$ and the filling factor is $\nu=N_p/L$). To characterize the possible phases and phase transitions in the system, we calculate the energies $E_n$ of the two lowest states, the Berry phase $\gamma$ and the fidelity metric $g$ of the ground-state. The Berry phase is defined as $\gamma =\oint i\langle \psi _{\theta }|\frac{d}{d\theta }|\psi _{\theta
}\rangle$ with $\theta$ the twisted boundary phase and its value $\gamma$ mod $2\pi$ gets a nonzero value $\pi$ for topological phase while zero for trivial phase \cite{berry1,berry2,berry3}. The fidility metric $g$ is defined as $g(V,\delta V)=\frac{2}{N}\frac{1-F(V,\delta V)}{(\delta V)^2}$ with the fidelity $F(V,\delta V)=|\langle \Psi_{0}(V)|\Psi_{0}(V+\delta V)$ the overlap of the two ground-state wave functions at $V$ and $V+\delta V$ \cite{int1}. When the topological band is filled with hard-core bosons, as $U$ is increased, it shows in Fig.\ref{fig1} that the ground-state remains gapped and $\gamma=0$, $g=0$ all the way, indicating that no obvious phase transition happens. This is in contrast to the fermionic case, where the Hubbard interaction $U$ drives a topological phase transition \cite{my1}.

\begin{figure}[tbp]
\includegraphics[width=8cm]{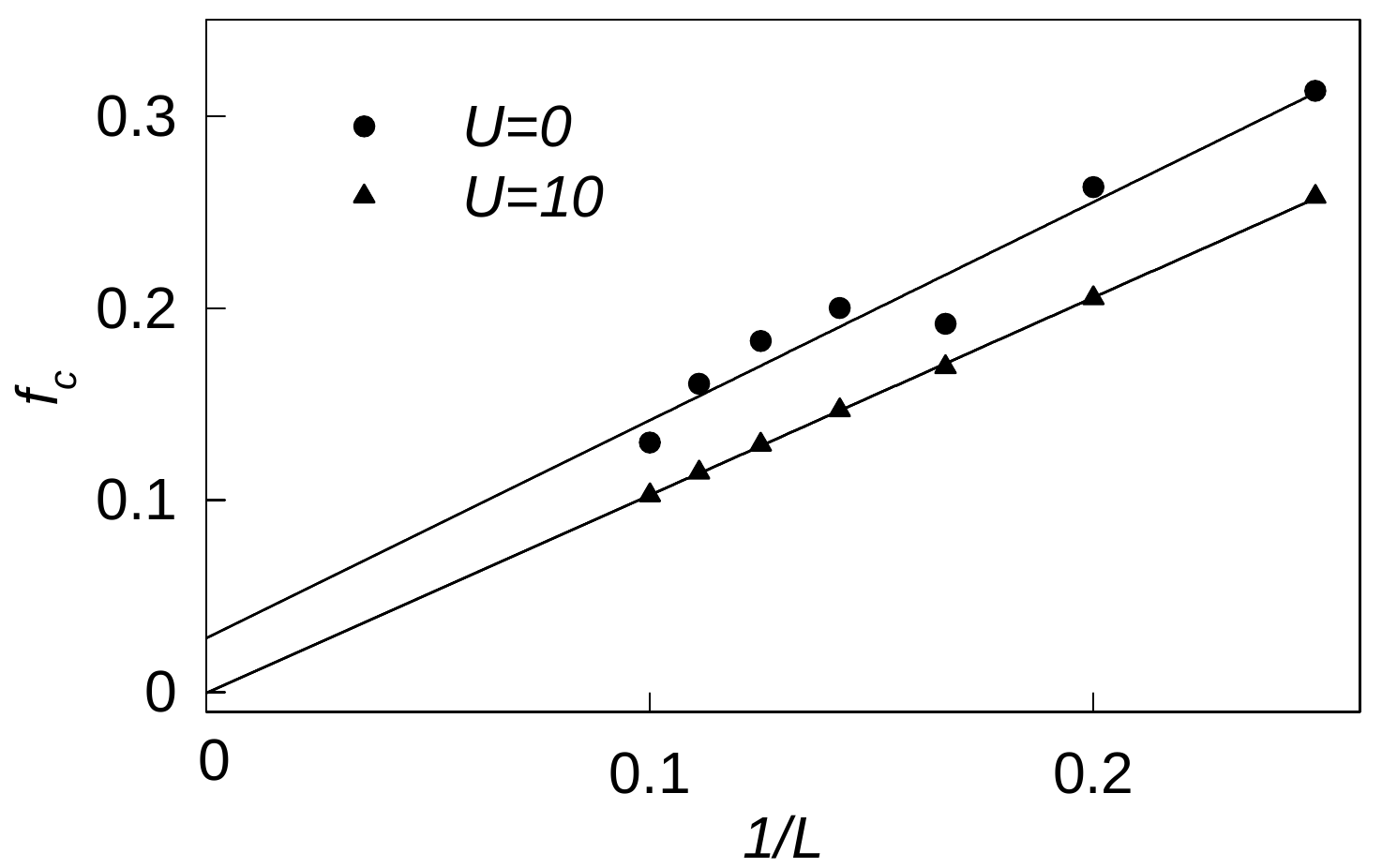}
\caption{Finite-size scaling of the condensate fraction $f_c$ at $U=0,10$. The parameters are the same with those in Fig.\ref{fig1}.}
\label{fig3}
\end{figure}

So the topological property  doesn't persist when hard-core bosons replace the fermions. To identify the bosonic phase, we study the momentum distribution, which is defined by the formula \cite{boson1}
\[
n_{L}(k)=\frac{1}{L}\sum_{i,j=0}^{L-1}\langle c_{i}^{\dagger}c_{j}+d_{i}^{\dagger}d_{j}\rangle e^{ik(i-j)},
\]%
with the momentum $k=(2\pi/L)l (l=0,1,...,L-1)$ and the average $\langle \cdots \rangle$ over the ground-state wave function. As has been known for free hopping bosons, the ground-state is a SF, which is characterized by the peak at zero-momentum state and its height strongly depending on $L$. In our case of $U=0$ (see Fig.\ref{fig2}), the momentum distribution shows peaks at $k=\pi/2,3\pi/2$ and theirs heights increase with the size $L$. Thus the results show that the system is in a trivial SF phase at $\nu=1$. When the Hubbard interaction is turned on, the system is expected to experience a phase transition to the Mott-insulator phase. It is from the result at $U=10$ [see Fig.\ref{fig2}(b)] where the momentum distribution $n_{L}(k)$ tends to be uniform and its values nearly don't change with $L$. To further characterize the phases, we measure the condensate fraction $f_c=(\Lambda_{c}+\Lambda_{d})/N_{b}$ ($N_{b}$ is the total number of the hard-core bosons) with $\Lambda_c$ ($\Lambda_d$) the largest eigenvalue of the one-particle density matrix $\rho_{ij}^{c}=\langle c_{i}^{\dagger}c_{j}\rangle (\rho_{ij}^{d}=\langle d_{i}^{\dagger}d_{j}\rangle)$ \cite{boson2}. In Fig.\ref{fig3} it shows that at $U=0$ $f_c$ scales to a nonzero value in the thermodynamic limit while at $U=10$ it scales to zero. So for small $U$ the system has a nonzero SF density, while for large $U$ the system becomes a Mott-insulator. We emphasize that the study of the SF to Mott-insulator transition needs scaling for systems with larger sizes, which is beyond the present method.

%

\section{The filling factor $\nu=1/3$}
\begin{figure}[tbp]
\includegraphics[width=7cm]{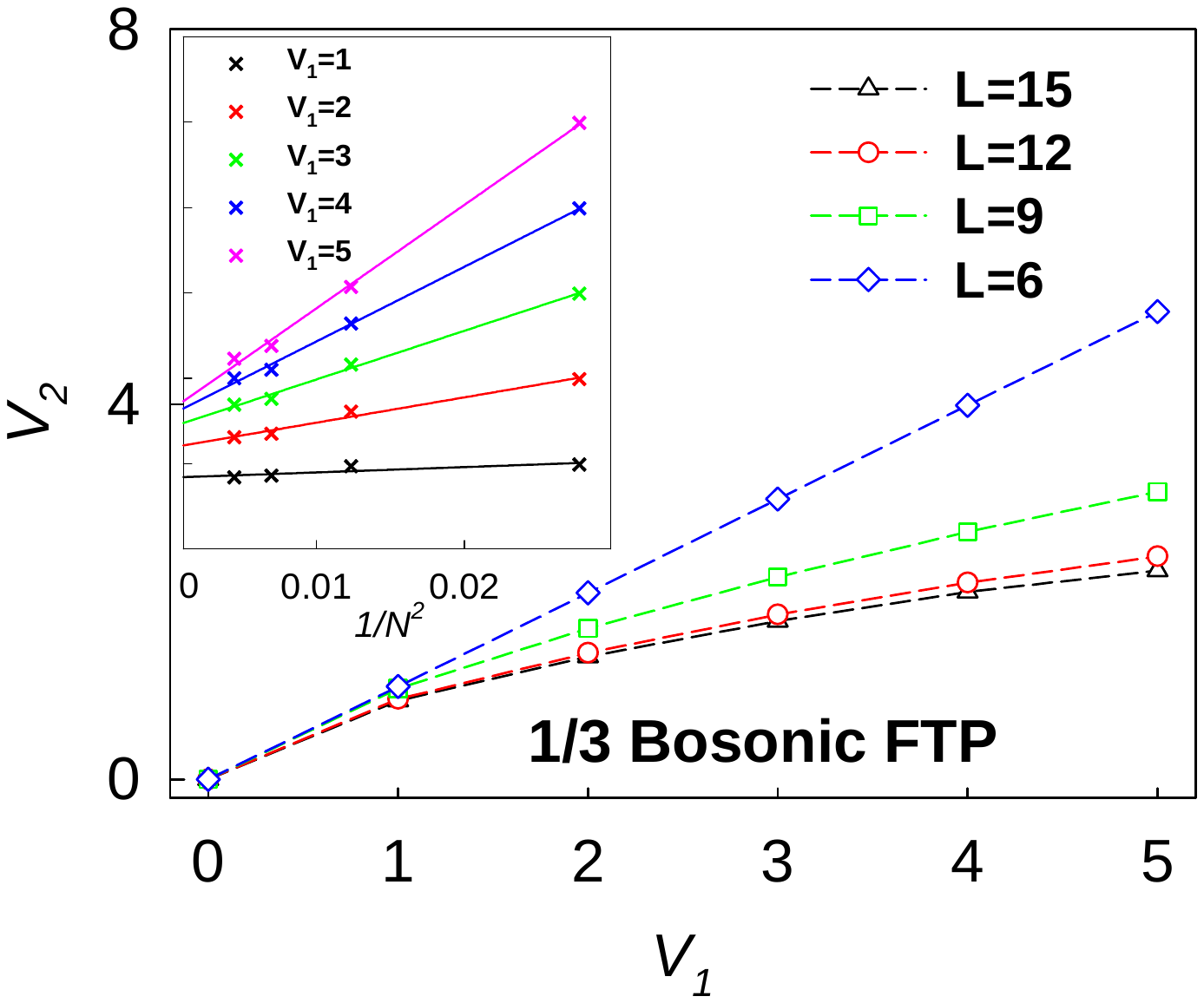}
\caption{(Color online) The phase diagram in the $(V_1,V_2)$ plane at $\nu=1/3$ for different sizes. The inset shows the finite-size scaling of the critical value $V_{2c}$ at different $V_1$. Here $A=B=1$ and $M=-2$ when the Hamiltonian Eq.(\ref{eq1}) has non-trivial flatband in the fermionic case. }
\label{fig4}
\end{figure}

Next we study the case of the filling factor $\nu=1/3$. We drop the Hubbard interaction, but add NN and NNN interactions to Eq.(\ref{eq1}), which writes,
\[
H_{I}=V_{1}\sum_{\langle i,j\rangle }n_{i}n_{j}+V_{2}\sum_{\langle \langle
i,j\rangle \rangle }n_{i}n_{j}
\]%
where $n_{i}=n_{i}^{c}+n_{i}^{d}$ is the total
number of hard-core bosons on site $\mathbf{r}_{i}$ and $V_{1}$, $V_{2}$
are the strength of the interactions. We have carried out
the calculations at $\nu=1/3$, and find the bosonic FTP where the ground-state is three-fold degenerate.
We first glance at the phase diagram in the $(V_{1},V_{2})$ plane,
which is shown is Fig.\ref{fig4}. By turning on $V_{1}$, the ground-state
is three-fold degenerate and the bosonic FTP emerges. The ground-state is separated
from higher eigenstates by a finite gap, whose value increases with the strength of $V_{1}$. After turning on $V_{2}$, the
value of the gap is decreased and vanishes at a critical value $V_{2c}$, which marks the boundary in the
phase diagram. Finite-size scaling shows that the bosonic FTP exists in the thermodynamic limit (see the inset of Fig.\ref{fig4}).

\begin{figure}[tbp]
\includegraphics[width=8cm]{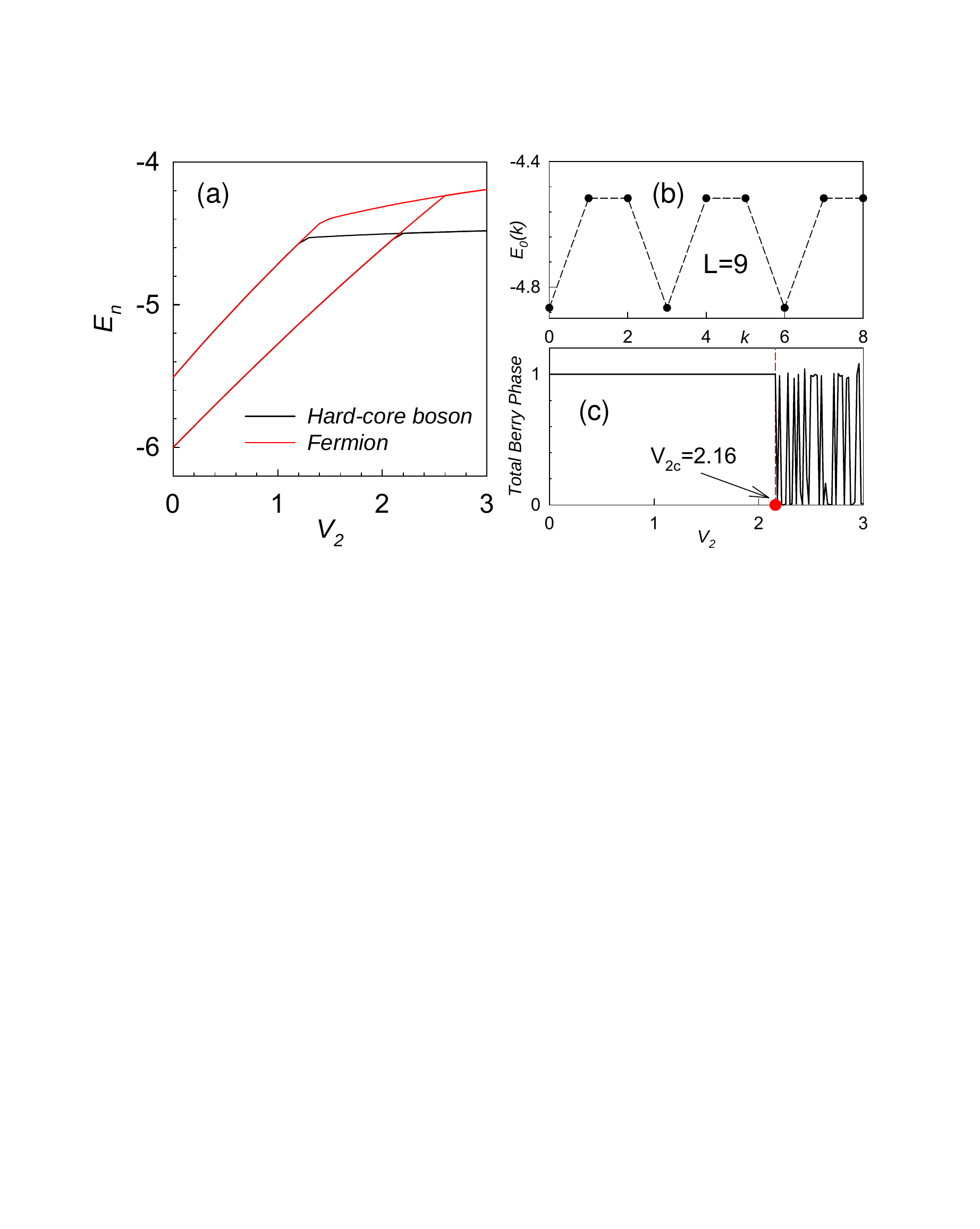}
\caption{(Color online)(a) The energies of the ground- and the first-excited states vs $V_{2}$ at $V_{1}=3$. (b) The ground-state energy of each momentum sector at $V_{1}=3$ and $V_2=1.6$. (c) The total Berry phase vs $V_2$ at $V_1=3$. The parameters are the same with those in Fig.\ref{fig4} except $M=-1.999$ in (c) when the band slightly departs the exact flatness. Here the system size is $L=9$.}
\label{fig5}
\end{figure}

We note that the present phase diagram is very similar to that of the corresponding fermionic system except the smaller critical values $V_{2c}$ \cite{my2}. In Fig.\ref{fig5}(a), it is shown more clearly: at small $V_2$ the energies $E_n$ of the two lowest states are almost the same in the two systems, while at larger $V_2$ they are different. The main difference of hard-core bosons and fermions is the exchanging relation. So when the number of the particles is fewer, the exchanging between the particles is less possible and the hard-core bosons are more like spinless fermions. Also at small $V_2$, there is one particle within each isolated bond, thus their difference is further weakened. It is interesting that the situation is similar in two dimensions where hard-core bosons in topological band at integer filling don't exhibit topological phase, while at fractional fillings they do \cite{int1,fqh4}.

In momentum space, the degenerate ground-states are in different momentum sectors and are equally spaced with the interval of $N_p$, as shown in Fig.\ref{fig5}(b). We also calculate the total Berry phase of the ground-states, which is shown in Fig.\ref{fig5}(c). It shows that the total Berry phase gets nontrivial value $\pi$ for small $V_2$ and begins to be random between $(0,\pi)$ from a critical value $V_{2c}$, at which the bosonic FTP is broken (here the randomness is due to the fact that the multi-fold degeneracy of the ground-states is greater than three). The obtained critical value $V_{2c}$ is in good consistent with that from the energy spectra.

\section{Mapping hard-core bosons to spinless fermions}
\begin{figure}[tbp]
\includegraphics[width=8cm]{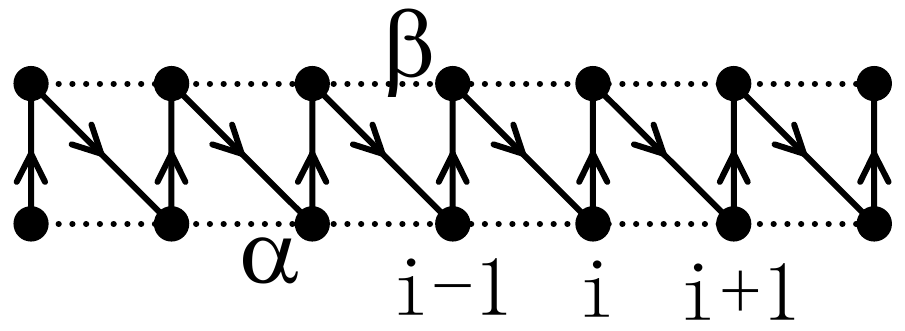}
\caption{Zigzag path in the two-leg ladder.}
\label{fig6}
\end{figure}

For 1D systems bosons and fermions can be transformed into each other and there has been example showing that the topological features can be manifested by Bose to Fermi statistics transmutations in other 1D systems \cite{segg}. So in the following we map hard-core bosons to spinless fermions using Matsubara-Matsuda and Jordan-Wigner transformations\cite{jordan1,jordan2,jordan3}, to gain some insights of the different behaviors of hard-core bosons at different filling factors from the mapped fermionic model. Using Matsubara-Matsuda transformation, the Hamiltonian Eq.(\ref{eq1}) can be mapped to a spin-$1/2$ one with the identification $c^{\dagger}_{i}(d^{\dagger}_{i})=S^{+}_{ic}(S^{+}_{id})$, $c_{i}(d_{i})=S^{-}_{ic}(S^{-}_{id})$ and $n_{ic,d}=\frac{1}{2}+S^{z}_{ic,d}$. The two-component system can be regarded as a two-leg ladder with one component on each site\cite{jordan4,jordan5}. For two-leg ladders Jordan-Wigner transformation can be applied directly when all sites are arranged in a 1D sequence (the zigzag path in Fig.\ref{fig6}). Then we divide the ladder into two sublattices and introduce two species of spinless fermions $\alpha_{i}$ and $\beta_{i}$. The spin operators on the two sublattices transform as:
\begin{eqnarray*}\label{eq2}
S^{+}_{i\alpha}&=&\alpha^{\dagger}_{i}e^{i\pi\sum_{j<i}(\alpha^{\dagger}_{j}\alpha_{j}+\beta^{\dagger}_{j}\beta_{j})} \\ \nonumber S^{+}_{i\beta}&=&\beta^{\dagger}_{i}e^{i\pi\sum_{j<i}(\alpha^{\dagger}_{j}\alpha_{j}+
\beta^{\dagger}_{j}\beta_{j})}e^{i\pi\alpha^{\dagger}_{i}\alpha_{i}}
\end{eqnarray*}
Using the above transformation, besides the terms in Eq.(\ref{eq1}) the following additional terms containing $4-$ and $6-$ fermion operators appear:
\begin{eqnarray}\label{eq3}
\Delta H=2B \sum_{i}\alpha^{\dagger}_{i}\alpha_{i+1}n_{i\beta}
-2B \sum_{i}\beta^{\dagger}_{i}\beta_{i+1}n_{i+1\alpha}\\ \nonumber
+2iA\sum_{i}\alpha^{\dagger}_{i}\beta_{i+1}
(n_{i\beta}+n_{i+1\alpha}-2n_{i+1\alpha}n_{i\beta})+H.c.
\end{eqnarray}
with $n_{i\alpha}=\alpha^{\dagger}_{i}\alpha_{i}$ and $n_{i\beta}=\beta^{\dagger}_{i}\beta_{i}$. At low fillings when there is no doubly occupying and neighboring, these additional terms vanish, so hard-core bosons show the same behaviors with fermions. While at high fillings, these terms show their effect and answer for the absence of the topological properties at $\nu=1$ in hard-core boson systems.

\section{Conclusions}
We have studied the hard-core bosons in 1D interacting topological bands at different filling factors. For the case of $\nu=1$, we consider on-site Hubbard interaction. By calculating the energies of the lowest states, the Berry phase and the fidelity metric of the ground-states, we find no sign of the existence of the bosonic topological phase, which is in contrast to the fermionic case. To identify the phase of the ground-state, we further study the momentum distribution and the condensate fraction and find a SF to Mott-insulator transition driven by the Hubbard interaction. For the case of $\nu=1/3$, we add NN and NNN interaction instead. We identify the bosonic FTP whose ground-states are characterized by the three-fold degeneracy and quantized total Berry phase. We also find that the obtained phase diagram is very similar to that of the corresponding fermionic system except the different critical values. Finally we discuss the reason of the different behaviors of hard-core bosons at different filling factors. Though the model we study is artificial, due to the rapid development of the field of cold-atoms  \cite{cold1}, it is hopeful that the model is engineered and the phases it exhibits are studied experimentally.
\section{Acknowledgements}
The author would like to thank Shiping Feng, Liu Bin, Jihong Qin and Shun-Qing Shen for helpful discussions. Specially the author thanks the Referee for valuable suggestions. Support for this work came from NSFC under Grant Nos. 11274032, 11104189 and FOK YING TUNG EDUCATION FOUNDATION.

\end{document}